# Mastermind is NP-Complete

Jeff Stuckman and Guo-Qiang Zhang [1]

*Department of Electrical Engineering and Computer Science*
*Case Western Reserve University*
*Cleveland, OH 44106*
*Email: gqz@eecs.case.edu*
*Web: newton.case.edu*

## 1 Introduction

In this paper we show that the Mastermind Satisfiability Problem (MSP) is NP-complete. The Mastermind is a popular game which can be turned into a logical puzzle called Mastermind Satisfiability Problem in a similar spirit to the Minesweeper puzzle [5]. By proving that MSP is NP-complete, we reveal its intrinsic computational property that makes it challenging and interesting. This serves as an addition to our knowledge about a host of other puzzles, such as Minesweeper [5], Mah-Jongg [1], and the 15-puzzle [6].

## 2 The Mastermind Game

The goal of Mastermind is for the player to determine the colors of each peg in a sequence of concealed locations (the solution). We first formalize the rules of Mastermind and then describe a variant called the Mastermind Satisfiability Problem which will be shown to be NP-complete.

In Mastermind, a player makes a series of guesses and receives responses to each guess as a rating of how close the guesses are to the solution. A player typically takes advantage of the feedback for previous guesses in order to inform the next guess, or determine the solution. A rating, or response, consists of the number of pegs in the guess having the *same color and position* as the corresponding peg in the solution and the number of pegs in the guess having *the same color but a different position* from a peg in the solution.

---

[1] Corresponding author.

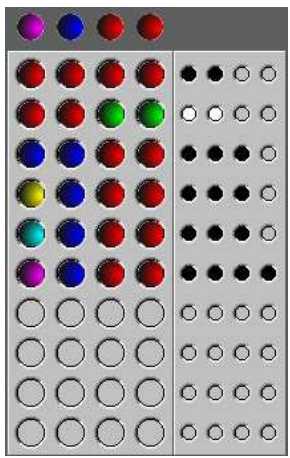

Fig. 1. A configuration of Mastermind with the key hidden in the top row.

Clearly, two parameters determine a specific Mastermind game, for the sake of formalization: one is the number $\kappa$ of colors, and the other is the length $\ell$ of the solution sequence. The number of guesses can be unbounded, although for each parameter pair $(\kappa, \ell)$, only a finite number of possible guesses exist without repetition. We use an initial segment of natural numbers $N_\kappa := [1, 2, \ldots, \kappa]$ to represent the colors, and an $\ell$-tuple in $N_\kappa^\ell$ to represent a guess.

In order to formulate solutions to the Mastermind game, we introduce a measure between two tuples to formalize the feedback information.

**Definition 2.1** *Let $x, y \in N_\kappa^\ell$. The Mastermind score between $x, y$ is defined as a pair of integers $\rho(x, y) := (b, w - b)$, where*

$$b := \#\{i \mid \exists i \in N_\ell,\ x_i = y_i\}$$

$$w := \sum_{j \in N_\kappa} \min(\#\{i \mid \exists i \in N_\ell,\ x_i = j\}, \#\{i \mid \exists i \in N_\ell,\ y_i = j\}).$$

*Here we use $\#A$ to denote the number of elements of a set $A$, and $x_i$ for the $i$-th element of a tuple $x$.*

If we think of $x$ as a guess and $y$ as the hidden solution, then $b$ captures the number of black pegs and $w - b$ captures the number of white pegs as a response for $x$. To see the latter, note that the value

$$\min(\#\{i \mid \exists i \in N_\ell,\ x_i = j\}, \#\{i \mid \exists i \in N_\ell,\ y_i = j\})$$

represents the total number of matches for a selected color $j$ between $x$ and $y$, in spite of the positions of the pegs. Thus summing this value over all possible colors and subtracting the number of pegs with both correct color and position result in the number of pegs with correct color but wrong position.

It is interesting to view the Mastermind score as the residuals of two distance measures. One is similar to the city-block distance, and the other is a distance based on the symmetric difference of multisets.

**Proposition 2.1** *For any $x, y \in N_\kappa^\ell$ define $\rho_1(x,y) := \ell - b$ and $\rho_2(x,y) := \ell - w$, where for the latter we regard each vector in $N_\kappa^\ell$ as an $\ell$-multiset for which repetition of elements is accounted for but not the order in which an element appears. Then $\rho_1$ and $\rho_2$ are distances in their respective spaces.*

**Proof.** In both cases only the triangular inequality needs to be checked; the symmetry and zero laws are trivial.

($\rho_1$). For $x, y, z$ in $N_\kappa^\ell$, the required triangular inequality $\rho_1(x,z) \le \rho_1(x,y) + \rho_1(y,z)$ translates to the inequality

$$\#\{j \mid \exists j \in N_\ell, \ x_j = y_j\} + \#\{k \mid \exists k \in N_\ell, \ y_k = z_k\} \le \ell + \#\{i \mid \exists i \in N_\ell, \ x_i = z_i\}.$$

For each $1 \le i \le \ell$, if $x_i = z_i$ then $\#\{i \mid x_i = y_i\} + \#\{i \mid y_i = z_i\} \le 2$, which can be rewritten as $\#\{i \mid x_i = y_i\} + \#\{i \mid y_i = z_i\} \le 1 + \#\{i \mid x_i = z_i\}$. If $x_i \ne z_i$, then $\#\{i \mid x_i = y_i\} + \#\{i \mid y_i = z_i\} \le 1$ no matter which value $y_i$ assumes. This can again be rewritten as $\#\{i \mid x_i = y_i\} + \#\{i \mid y_i = z_i\} \le 1 + \#\{i \mid x_i = z_i\}$. Summing up over all $i$ in the range $[1, \ell]$, the desired inequality follows.

($\rho_2$). Let $[x], [y], [z]$ be elements of $[N_\kappa^\ell]$, where $[\ ]$ stands for the projection of vectors in $N_\kappa^\ell$ as multisets (e.g. $[(1,3,3,1)] = \{1,1,3,3\}$, and $[(1,3,3,1)] = [(3,1,1,3)]$). Then $\rho_2([x],[y]) = \#([x]-[y]) + \#([y]-[x])$, the size of the symmetric difference of multisets. It is straightforward to check that the size of symmetric difference is indeed a distance measure. □

The realization that the Mastermind score consists of two independent distance measures provides a basis for computer implementation as a search problem in high dimensional spaces.

## 2.1 The Mastermind Satisfiability Problem

A Mastermind variant is the Static Mastermind [4], for which the guesses are all given at once to receives a collective response. The player then tries to figure out the solution. Goddard [3] provides a combinatorial analysis of upper bound on the number of guesses for small configurations in order to deduce a solution.

We approach Mastermind as a decision problem: given a set of guesses $G \subseteq N_\kappa^\ell$ and their corresponding scores, is there at least one valid solution? We refer to this problem as the Mastermind Satisfiability Problem (MSP) and show that it is NP-complete with respect to size $\ell$ (for $\kappa > 1$).

Here is a formal statement of MSP.
**Input**: $G$, a subset of $N_\kappa^\ell$ and for each $g \in G$, a Mastermind score $(b_g, w_g)$.
**Output**: YES if there exists an element $s \in N_\kappa^\ell$ such that for each $g \in G$, $\rho(g, s) = (b_g, w_g)$, and NO otherwise.

Our main result of this section is the following.

**Theorem 2.1** *MSP is NP-complete.*

**Proof.** It is apparent that the validity of a solution for an instance of MSP can be evaluated in polynomial time, because checking a satisfying peg configuration is as easy as matching the pegs against each guess.

We show that MSP is NP-hard by reducing the NP-hard Vertex-Cover Problem (page 1006, [2]; see also [7]) to it. The Vertex-Cover($n$) Problem is to determine if there exists a size-$n$ subset of vertices of a graph such that each edge borders at least one vertex in the selected subset.

We translate an instance of Vertex-Cover($n$) Problem to an instance of MSP. Let $G = (V, E)$ be a graph, and $n > 1$. For its corresponding MSP instance, set $\kappa = \#V + \#E + 2$ and $\ell = 3 + 2\#V + \#E$. The idea is to encode each vertex and each edge of $G$ as a distinct color, plus two control colors. The parameter $\ell$ makes room for the first three positions for encoding edges, the next $2\#V$ positions for vertex selection, and the last $\#E$ positions for edge selection. $2\#V$ positions are reserved for vertex selection to make sure that there is no location overlap between a vertex in the guess and a vertex in the solution. For convenience, the colors are labeled explicitly, as follows:

$$K = \{v_1, v_2, v_3 \ldots v_{\#V}, e_1, e_2, e_3 \ldots e_{\#E}, Y, N\}$$

Now the set of guesses can be given as follows:

(1) The first guess will be $(N, N, \ldots, N)$ with a score of $(0, 0)$ to prevent the control element $N$ from appearing in the solution.
(2) The second guess will be $(Y, Y, Y, N, \ldots, N)$ with a score of $(3, 0)$ to force the first three elements of the solution to be the control element $Y$.
(3) Next, create one row for each edge of the graph. For the $i$-th edge $(a, b)$ in $E$, create the guess $(e_i, a, b, N, N, \ldots, N)$ with a score of $(0, 2)$. Note that because of the previous item, the first three positions are cleared from being part of the solution and thus there is no correct position for this guess.
(4) Finally, create a guess $(Y, Y, Y, v_1, v_2, v_3, \ldots, v_{\#V}, N, N, \ldots, N)$ with a score of $(3, n)$. Note that this score accounts for the number of correct colors from $V$ but leaves their positions unspecified.

We show that the Vertex-Cover Problem $(G, n)$ has a solution if and only if the instance of MSP above has a solution.

(*If*). Suppose the MSP instance described above has a solution. By constraint (1), color $N$ does not appear in the solution. Hence, precisely $n$ vertices $w_1, w_2, \ldots, w_n$ from $V$ appear as part of the solution, as specified in constraint (4). We verify that for each edge in $E$, at least one vertex in $W$ is adjacent to it. Note that each edge has a corresponding guess in constraint (3). Since the solution must satisfy this constraint, at least one vertex among $a$ and $b$ must be in $W$ to receive a score $(0, 2)$. Therefore $W$ is a size-$n$ vertex-cover for $G$.

(*Only If*). Suppose $W = \{w_1, w_2, \ldots, w_n\}$ is a size-$n$ vertex-cover for $G$. Then the corresponding MSP solution can be given as

$$(Y, Y, Y; Y, \ldots, Y; w_1, w_2, \ldots, w_n, Y, \ldots, Y; e_{i_1}, e_{i_2}, \ldots, e_{i_t}, Y, \ldots, Y),$$

where $e_{i_j} = (a, b)$ appears in this solution if and only if $\{a, b\} \not\subseteq W$, i.e., $e_{i_j}$ is an edge using precisely one vertex in $W$. Here we used semicolon ; to clearly indicate distinct regions: the first region with three positions are reserved for edges, the next region of $\#V$ positions are reserved for guesses, and the $\#V$ positions after are where the selected edge set is located. We need to place precisely those edges with precisely one vertex in the selected vertex set in the solution, without letting any edge with both vertices in the solution to appear, so that the score $(0, 2)$ for edges comes out right. The remaining $Y$s are used as padding. With these in mind, it is quite straightforward to check that the scores are correct for all the guesses described in (1) - (4) with respect to this specific Mastermind solution.

It is also clear that our reduction is polynomial in input size. □

*Remark.* We used $\ell = 3 + 2\#V + \#E$ in the proof to make it more crisp; this parameter can be reduced to $3 + \#V + \#E$ by taking advantage of vertex permutations. Given a specific arrangement of vertices $(v_1, v_2, v_3, \ldots, v_{\#V})$, there exists a permutation $(u_1, u_2, \ldots, u_{\#V})$ with $u_i \in W \cup \{Y\}$ for each $i = 1, \ldots, \#V$ such that $v_i \neq w_i$ for all $1 \leq i \leq \#V$, and each element of $W$ appears exactly once in the permutation (assuming either $n \neq \#V$, or $n > 1$). This permutation can then be used in constructing a solution for the "Only If" part above.

## 3 Uniqueness of Solution

In logical puzzles, the most interesting cases are those configurations for which the solution is unique. Determining MSP instances with unique solutions is no more complex than finding the solutions in general, in the sense that any algorithm for finding a Static Mastermind solution can be turned into an algorithm to determine if the solution is unique.

Assume that we have an algorithm that finds a solution $s = (s_1, s_2, s_3, \ldots)$ for a Static Mastermind instance $G$. Add $s$ as a guess and create an instance of Static Mastermind *for each pair* $(b, w) \neq (\ell, 0)$ where $(b, w)$ is the score for $s$. Then the solution $s$ to the original input $G$ is unique if and only if one of the new instances with score $(b, w) \neq (\ell, 0)$ has a solution. The total number of score pairs $(b, w)$ other than $(\ell, 0)$ is

$$\sum_{b=0}^{\ell-1}(\ell - b + 1) = \left(\sum_{b=0}^{\ell-1} -b\right) + \ell^2 + \ell = \frac{\ell \cdot (\ell + 3)}{2}.$$

Because the number of times that the Static Mastermind algorithm must be run (on slightly modified instances) is at most polynomial, verifying that a solution is unique is no harder than finding the solution itself, assuming that finding the solution itself was polynomial or harder.